\documentclass[english,aps,prl,twocolumn,amsmath,amssymb,showpacs,superscriptaddress,notitlepage,longbibliography]{revtex4-2}
\usepackage{bm}
\usepackage[pdftex]{graphicx,hyperref}
\hypersetup{colorlinks = true, urlcolor = blue, linkcolor = blue, citecolor = blue}
\usepackage{mathtools}
\usepackage{ulem}
\usepackage{color}
\usepackage{bbding}

\begin{document}
	
	\title{Spin-Orbital Altermagnetism}
	
	\author{Zi-Ming Wang}
	\affiliation{Department of Physics and Chongqing Key Laboratory for Strongly Coupled Physics, Chongqing University, Chongqing 400044,  China}
	\affiliation{Center for Correlated Matter and School of Physics, Zhejiang University, Hangzhou 310058, China}
	
	\author{Yang Zhang}
	\email{yzhang@utk.edu}
	\affiliation{Department of Physics and Astronomy, University of Tennessee, Knoxville, Tennessee 37996, USA}
	
	\author{Song-Bo Zhang}
	\email{songbozhang@ustc.edu.cn}
	\address{Hefei National Laboratory, Hefei, Anhui, 230088, China}
	\address{International Center for Quantum Design of Functional Materials (ICQD), University of Science and Technology of China, Hefei, Anhui 230026, China}
	
	\author{Jin-Hua Sun}
	\affiliation{Department of Physics, Ningbo University, Ningbo 315211, China}
	
	\author{Elbio Dagotto}
	\affiliation{Department of Physics and Astronomy, University of Tennessee, Knoxville, Tennessee 37996, USA}
	
	\author{Dong-Hui Xu}
	\email{donghuixu@cqu.edu.cn}
	\affiliation{Department of Physics and Chongqing Key Laboratory for Strongly Coupled Physics, Chongqing University, Chongqing 400044, China}
	\affiliation{Center of Quantum Materials and Devices, Chongqing University, Chongqing 400044, China}
	
	\author{Lun-Hui Hu}
	\email{lunhui@zju.edu.cn}
	\affiliation{Center for Correlated Matter and School of Physics, Zhejiang University, Hangzhou 310058, China}
	\affiliation{Department of Physics and Astronomy, University of Tennessee, Knoxville, Tennessee 37996, USA}

	\begin{abstract}
		Altermagnet is a newly discovered magnetic phase, characterized by non-relativistic spin-splitting that has been experimentally observed. Here, we introduce a framework dubbed {\it spin-orbital altermagnetism} to achieve spin-orbital textures in altermagnetic materials. We identify two distinct classes of spin-orbital altermagnetism: intrinsic and extrinsic. The intrinsic type emerges from symmetry-compensated magnetic orders with spontaneously broken parity-time symmetry, while the extrinsic type stems from translational-symmetry breaking between sublattices, as exemplified by the Jahn-Teller-driven structural phase transition. In addition to directly measuring the spin-orbital texture, we propose spin conductivity and spin-resolved orbital polarization as effective methods for detecting these altermagnets. Additionally, a symmetry-breaking mechanism induces weak spin magnetization, further revealing the peculiar feature of spin-orbital altermagnetism. We also utilize the staggered susceptibility to illustrate a potential realization of this phase in a two-orbital interacting system. Our work provides a new platform to explore spin-orbital locked physics, extending the materials classes that may display complex spin textures from the standard $4d-5d$ compounds to $3d$ compounds. 
	\end{abstract}
	
	\maketitle

	\textit{\color{blue}Introduction.--}
	Spin and orbit are two fundamental degrees of freedom of electrons, and their intertwining is crucial for understanding various electronic states and phenomena in quantum materials~\cite{tokura2000orbital}. Spin-orbit coupling (SOC), a relativistic effect in solids, couples these degrees of freedom, lifting Kramers spin degeneracy and resulting in spin-split energy bands with distinctive spin textures in reciprocal space. Recent experiments using spin-resolved and photon-polarized angle-resolved photoemission spectroscopy~\cite{sobota2021rmp} have revealed entangled spin-orbital textures—a locking phenomenon between spin and atomic orbital degrees of freedom—in strongly spin-orbit coupled systems such as topological insulators~\cite{cao2013np,zhang2013prl,zhu13prl}. This breakthrough opens novel avenues for manipulating spin polarization by targeting the orbital domain.

	Non-relativistic mechanisms for spin-splitting effects have also garnered significant attention in recent years. These SOC-free phenomena have become prominent after the discovery of altermagnetism (AM)~\cite{vsmejkal2020crystal,Naka19NC,Ahn19PRB,hayami2019momentum,Yuan2020Giant,mazin2021prediction,ma2021multifunctional,ifmmode2022Beyond,ifmmode2022Emerging,krempasky2024altermagnetic,sdongFeSb2}, which is a new magnetic phase characterized by momentum-dependent spin splitting despite zero net magnetization~\cite{jungwirth2024altermagnets,bai2024altermagnetism,fender2025altermagnetism,liu2025different}. Subsequent experimental studies have confirmed AM-induced spin-splitting bands in diverse materials~\cite{fedchenko2024ruo2,lin2024ruo2,gonzalez2023prl,krempasky2024MnTe,lee2024MnTe,osumi2024MnTe,liu2024chiral,reimers2024CrSb,ding2024CrSb,yang2025three,jiang2025metallic,zhang2025crystal}. A few earlier theoretical works for non-relativistic spin splitting were established through spin-channel Pomeranchuk instabilities~\cite{Hirsch1990prb,wu2004prl,CJWu07PRB} and $d$-wave spin-density wave states~\cite{Ikeda1998prl}. Furthermore, non-coplanar antiferromagnetic materials also exhibit significant non-relativistic spin-split bands~\cite{Chen2014Anomalous,nakatsuji2015large,vsmejkal2022anomalous,ren2023enumeration,jiang2023enumeration,xiao2023spin,liu2018giant,zhu2024observation}. These novel magnetic
	systems can exhibit intriguing properties and potential functionalities~\cite{Nakaprb,shao2021spin,FengZX22NE,Fernandestoplogical,zhang2024prl}.

	Although distinct mechanisms for AMs have been proposed recently~\cite{he2023prl,bhowal2024prx,leeb2024prl,das2024prl,atasi2024prb,vsmejkal2024altermagnetic,duan2025prl,gu2025ferroelectric,zhu2025two}, other spin-related non-relativistic phenomena, particularly the physics of spin-orbital textures without SOC, remain largely unexplored. This can be achieved through intertwined symmetry-compensated magnetic orders that couple these two degrees of freedom. Here, we establish a new theoretical framework to attain non-relativistic spin-orbital textures through altermagnetic orders, dubbed \textit{spin-orbit altermagnetism}. Among various altermagnetic orders permitted by spin-space symmetry~\cite{ifmmode2022Beyond}, we identify two distinct classes of spin-orbital altermagnetism: one arising intrinsically and the other necessitating crystalline symmetry breaking due to structural phase transitions. We demonstrate that the spin conductivity and spin-resolved orbital polarization are effective tools for detecting and distinguishing between these phases. We also examine the weak ferromagnetic magnetization caused by symmetry breaking, relevant to the hysteresis loop observed in the anomalous Hall effect. Furthermore, the staggered susceptibility is employed to illustrate a potential realization of such a phase in two-orbital interacting systems.

\begin{figure}[t]
		\centering
		\includegraphics[width=0.95\linewidth]{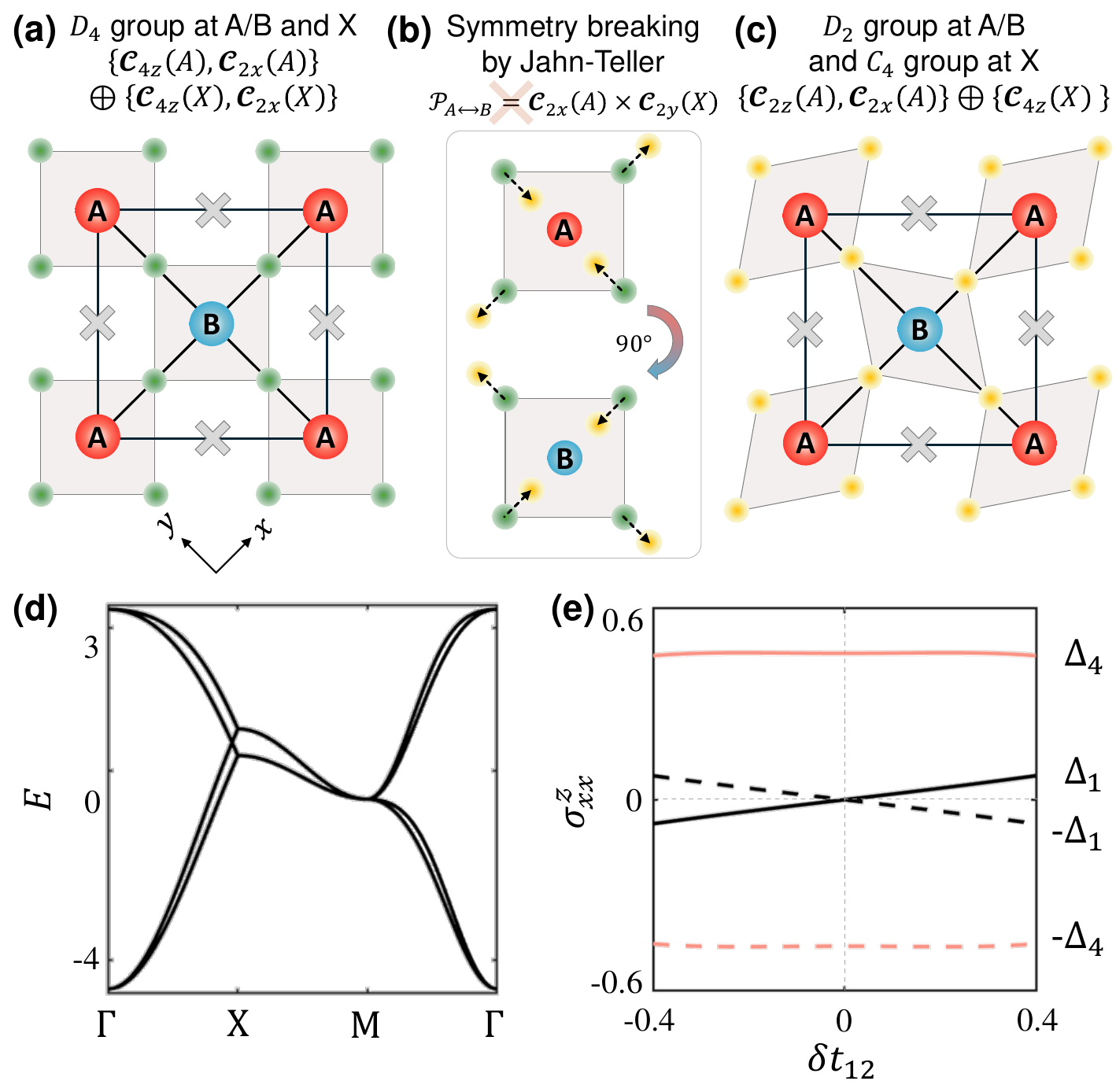}
		\caption{(a) The square lattice with A (red) and B (blue) sublattices, each surrounded by four symmetric atoms (green). 
			(b) Jahn-Teller distortions of green atoms around A and B, indicated by opposite-directional dot arrows.
			(c) The distorted lattice by combining (a) and (b). 
			(d) Tight-binding band structure for the lattice in (c). 
			(e) Spin conductivity $\sigma_{xx}^z$ as a function of JT distortion strength $\delta t_{12}$ for ${\cal O}_{1}$ (JT-induced) and ${\cal O}_{4}$ (intrinsic) phases. 
		}
		\label{fig1}
	\end{figure}

	\textit{\color{blue}Intrinsic altermagnet.--}
	We study a square lattice with two sublattices, A and B, surrounded by a crystal field generated by four symmetrically positioned atoms, as shown in Fig.~\ref{fig1}(a). If A is identical to B, the system exhibits the $D_4$ point group symmetry at both A and B sites and the centroid of the square (labeled ``X'' in gray). The symmetry generators are $\{ {\cal C}_{4z}(A), {\cal C}_{2x}(A) \}$ and $\{ {\cal C}_{4z}(X), {\cal C}_{2x}(X)\}$, leading to three symmetries that relate A to B: (i) four-fold rotation ${\cal C}_{4z}(X)$, (ii) inversion ${\cal P}_{A\leftrightarrow B} = {\cal C}_{2x}(A) {\cal C}_{2y}(X)$, and (iii) mirror ${\cal M}_{A\leftrightarrow B} = {\cal C}_{4z}(X) {\cal C}_{2x}(A)$. In this case, the symmetry-compensated magnetic order arises from breaking time-reversal symmetry ${\cal T}$ while preserving at least one of the following spin-space group symmetries: $[{\cal C}_2||{\cal P}_{A\leftrightarrow B}]$, $[{\cal C}_2||{\cal C}_{4z}(X)]$ and $[{\cal C}_2||{\cal M}_{A\leftrightarrow B}]$~\footnote{In the Sec.~I of the SM~\cite{sm2024}, we additionally employ conventional magnetic space group notation to represent these symmetries: ${\cal P}_{A\leftrightarrow B}{\cal T}$, ${\cal C}_{4z}(X){\cal T}$ and ${\cal M}_{A\leftrightarrow B}{\cal T}$. Importantly, in the absence of SOC, these symmetry operations maintain identical physical interpretations to the spin-space group symmetries.}. Here, the left element ${\cal C}_2$ acts solely on spin space (i.e.,~flipping spins), while the right elements act purely in real space. Since SOC is absent or negligible, the spin-space group provides a comprehensive framework for classifying collinear magnetic orders~\cite{ifmmode2022Beyond} and others~\cite{liu2022prx}. Furthermore, in a two-orbital system, two additional symmetries, $[{\cal C}_2||{\cal C}_{4z}(A)]$ and $[{\cal C}_2||{\cal C}_{2x}(A)]$, permit a staggered magnetic order between the two orbitals. Thus, based on spin-space symmetry analysis (see Sec.~I of Supplementary Material (SM)~\cite{sm2024}), five staggered orders are identified in Table~\ref{table1}. These orders all feature zero net magnetization and are generally classified as antiferromagnetic or altermagnetic. Their matrix representations ${\cal O}_{1\to5}$ are $\{ \tau_0\sigma_zs_z,  \tau_x\sigma_zs_z, \tau_x\sigma_0s_z, \tau_z\sigma_0s_z, \tau_z\sigma_zs_z \}$, where $\bm{\tau},\bm{\sigma}$, and $\bm{s}$ denote Pauli matrices for orbital, sublattice, and spin degrees of freedom, respectively.

	\begin{table}[b].
		
		\centering
		\caption{Symmetry classification and properties of five staggered orders, with or without inversion (${\cal P}_{A\leftrightarrow B}$), in a two-orbital square-lattice system with sublattices A and B. The spin-space group operator $g$ acts on ${\cal O}_i$ as follows: $g[{\cal O}_i]g^\dagger = \chi_i {\cal O}_i$, where the character $\chi_i=\pm 1$ corresponds to the symbols \CheckmarkBold and \XSolidBrush, respectively. Here, AM, AFM, FM, and SO denote altermagnetism, antiferromagnetic, ferromagnetism, and spin-orbital, respectively. $g$-wave is planar type.
		}
		\includegraphics[width=\linewidth]{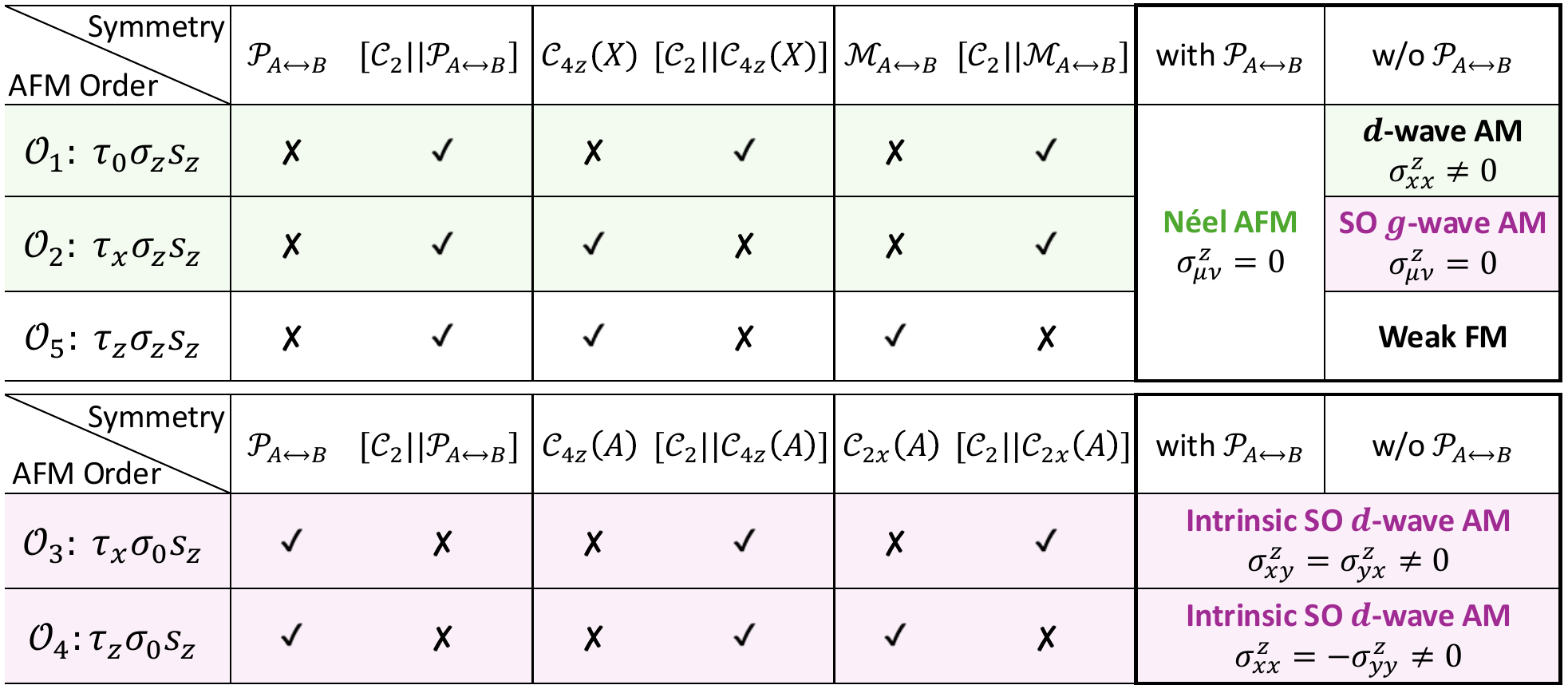}
		\label{table1}
	\end{table}

	We now discuss general results in Table~\ref{table1} with ${\cal P}_{A\leftrightarrow B}$. Collinear antiferromagnets can be either N\'eel or altermagnetic type~\cite{ifmmode2022Beyond}. The ${\cal O}_{1}$ phase represents the N\'eel order, which preserves ${\cal P}_{A\leftrightarrow B}{\cal T}$, similar to ${\cal O}_{2}$ and ${\cal O}_{5}$. According to Kramers' theorem, bands in these phases maintain spin degeneracy. However, ${\cal O}_{3}$ and ${\cal O}_{4}$ violate ${\cal P}_{A\leftrightarrow B}{\cal T}$ spontaneously, qualifying them as intrinsic AMs. They both feature $d$-wave altermagnetism, with subtle differences. For ${\cal O}_{3}$, it adheres to $[{\cal C}_2||{\cal C}_{2x}(A)]$, which enforces the band constraint $[{\cal C}_2||{\cal C}_{2x}(A)] \epsilon_n(s,k_x,k_y) = \epsilon_n(-s,k_x,-k_y)$, leading to nodal lines along $k_x=0$ or $k_y=0$. In contrast, the ${\cal O}_{4}$ phase shows $[{\cal C}_2||{\cal M}_{A\leftrightarrow B}]$-protected nodal lines along $k_x=\pm k_y$.

	\textit{\color{blue}Extrinsic altermagnet.--}
	In addition to ${\cal O}_{3/4}$, other symmetry-compensated orders can transition into AMs via symmetry breaking. Here, we examine crystalline symmetry breaking induced by the Jahn-Teller (JT) mode [Fig.~\ref{fig1}(b)]. In this case, the directions of lattice distortions for the A and B sublattices are opposite, resulting in a different crystal field environment compared to the symmetric case [Fig.~\ref{fig1}(a)]. As shown in Fig.~\ref{fig1}(c), the symmetry of the distorted lattice reduces to the $D_2$ group at A/B ($\{ {\cal C}_{2z}(A), {\cal C}_{2x}(A)\}$) and the $C_4$ group at X ($\{ {\cal C}_{4z}(X)\}$). Specifically, the JT mode breaks the ${\cal P}_{A\leftrightarrow B}$ symmetry or equivalently $C_{2y}(X)$.

	Table~\ref{table1} summarizes the effects of breaking ${\cal P}_{A\leftrightarrow B}$ symmetry while preserving other symmetries. Here we focus on the extrinsic spin-splitting bands in phases ${\cal O}_{1/2/5}$. By symmetry, ${\cal O}_{1}$ belongs to a $d$-wave AM with nodal lines along $k_x=\pm k_y$. ${\cal O}_{2}$ is a planar $g$-wave spin-orbital AM, displaying four nodal lines along $k_x=0$, $k_y=0$ and $k_x=\pm k_y$. As expected, ${\cal O}_{5}$ breaks all the symmetries required to classify it as an antiferromagnetic phase, so that it exhibits a non-zero net spin magnetization.

	\begin{figure}[t]
		\centering
		\includegraphics[width=\linewidth]{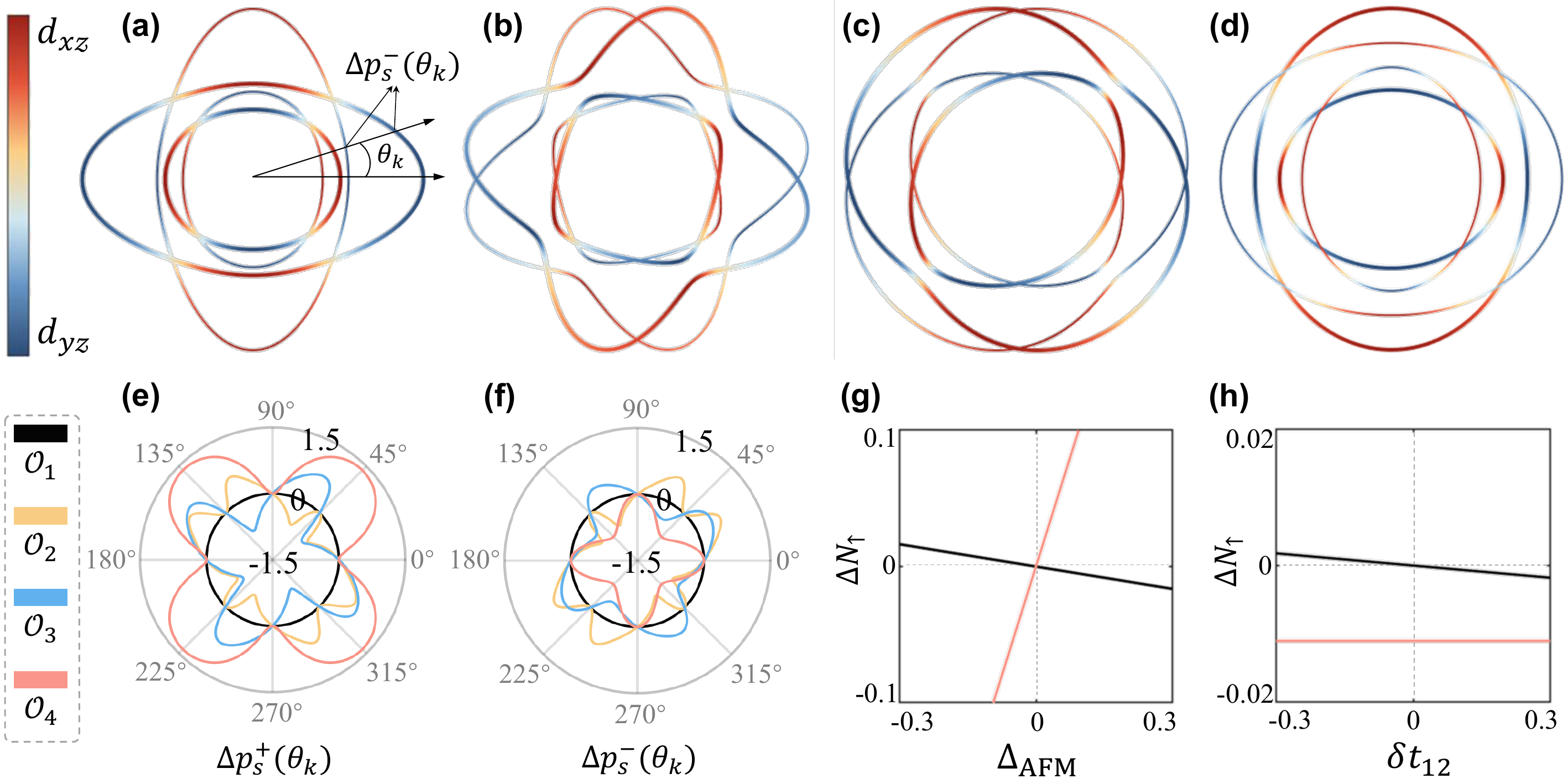}
		\caption{(a-d) Spin-split Fermi surfaces for ${\cal O}_{1/2/3/4}$, respectively, with the thick (thin) lines denoting the $s_z=\uparrow$ ($\downarrow$) bands. The color scheme represents the orbital weight on the Fermi surfaces.
			(e,f) Corresponding spin-orbital textures $\Delta p_s^\pm(\theta_k)$ for the $+$ (inner Fermi surfaces) and $-$ bands (outer Fermi surfaces), respectively, as illustrated in (a). 
			(g,h) Spin-resolved orbital polarization as a function of the altermagnetic gap $\Delta_{\text{AFM}}=\Delta_{1}$ (black) and $\Delta_{\text{AFM}}=\Delta_{4}$ (red), and JT distortion strength $\delta t_{12}$, respectively. 
			Parameters for the ${\bm k}\cdot{\bm p}$ model are derived from the tight-binding model used in Fig.~\ref{fig1}(d),  with $m_0=-0.2$ for better visualization. Other parameters are $\mu=2.5$, $\Delta_{1}=2.5$, $\Delta_{2}=3.5$, and $\Delta_{3/4}=0.2$.
		}
		\label{fig2}
	\end{figure}

	To quantitatively investigate the influence of the JT mode, we employ a tight-binding model within the mean-field formalism for the staggered orders,
	\begin{align} \label{eq-mf-ham0}
		\begin{split}
			{\cal H}_\text{MF}({\bm k}) = & \;{\cal H}_0({\bm k})s_0 + \Delta_1 \tau_0 \sigma_z s_z + \Delta_2 \tau_x\sigma_z s_z \\
			&+\Delta_3 \tau_x \sigma_0 s_z + \Delta_4 \tau_z \sigma_0 s_z + \Delta_5 \tau_z\sigma_z s_z,
		\end{split}
	\end{align}
	where $\Delta_i$ with $i\in\{1,...,5\}$ are order parameters corresponding to the ${\cal O}_{i}$ orders, and we assume that $s_z$ is conserved. The non-interacting part is ${\cal H}_0({\bm k})= \tau_0 [ \epsilon_{0}({\bm k}) \sigma_0  +\epsilon_{1}({\bm k}) \sigma_x  + \epsilon_{3}({\bm k}) \sigma_z ] + [f_1({\bm k}) \tau_x + f_3({\bm k}) \tau_z ]\sigma_0$ with $ \epsilon_{0}({\bm k}) = - (t_1+t_2) (\cos{2 k_x} + \cos{2 k_y})$, $\epsilon_{1}({\bm k}) = -2 t_0 (\cos{k_x}+\cos{k_y})$, $\epsilon_{3}({\bm k}) = (t_2-t_1) (\cos{2 k_x} - \cos{2 k_y})$, {$f_1({\bm k}) = - 4t_3 \sin{k_x} \sin{k_y}$, and $f_3({\bm k}) = 2 t_5 (\cos{2 k_x} - \cos{2 k_y)}$. Cartoon representations of these hoppings are illustrated in Sec.~I of SM~\cite{sm2024}. We consider the inter-sublattice hopping $t_0$ as the dominant energy scale, which has been demonstrated to support the metallic AM phase~\cite{maierPRB2023weak}. Moreover, we define $\delta t_{12} \equiv (t_1-t_2)/(t_1+t_2)$ as a measure of the lattice distortion while disregarding other factors. The band structure for the distorted lattice is presented in Fig.~\ref{fig1}(d), using the parameters $t_0=1$, $t_3=0.07$, $t_5=0.04$, $t_1=0.2$, and $t_2=0.1$ (i.e.,~$\delta t_{12}=1/3$). In our JT-induced extrinsic altermagnetism, $\delta t_{12}\neq 0$ resembles the effects of lattice strain effect and ferroelectricicty~\cite{atasi2024prb,vsmejkal2024altermagnetic,duan2025prl,gu2025ferroelectric,zhu2025two}, both of which also break ${\cal P}_{A\leftrightarrow B}$.

		The JT-induced AM phases can be detected via spin-conductivity measurements~\cite{gonz2021prl,ifmmode2022Giant,bose2022tilted}. The spin current carries spin polarization along the $z$ direction, given by $J^z_\mu = \sigma_{\mu\nu}^z E_{\nu}$ with $\mu,\nu\in\{x,y\}$. Symmetry constraints dictate that the nonzero components are $\sigma_{xx}^z=-\sigma_{yy}^z\neq0$ for ${\cal O}_{1}$ and ${\cal O}_{4}$, $\sigma_{xy}^z = \sigma_{yx}^z\neq0$ for ${\cal O}_{3}$, whereas all components vanish for ${\cal O}_{2}$. We calculate explicitly $\sigma_{\mu\nu}^z$ as a function of $\delta t_{12}$ for ${\cal O}_{1}$ (extrinsic AM) and ${\cal O}_{4}$ (intrinsic AM), employing the linear response theory~\cite{mahan2013many}. The results are shown in Fig.~\ref{fig1}(e). $\sigma_{xx}^z$ changes sign upon reversing $\Delta_{1}$ or $\Delta_4$. Below the N\'eel temperature ($T<T_N$), the system can exhibit $d$-wave or $g$-wave AM, following a structural phase transition ($T<T_S$). In the $d$-wave scenario, if $T_N>T_S$, $\sigma_{xx}^z\neq0$ in the ${\cal O}_{1}$ phase only when $T<T_S$. Thus, spin conductivity can be a new means to detect the JT mode, serving as an alternative to scanning transmission electron microscopy~\cite{kim2023geometric}. While $\sigma_{xx}^z$ is small for ${\cal O}_1$ at small $\delta t_{12}$, it can be enhanced by tuning $\mu$ or considering JT-induced crystal fields (see Sec.~II of SM~\cite{sm2024}). Notably, $\sigma_{xx}^z$ of extrinsic AMs undergoes a sign change when the sign of $\delta t_{12}$ is inverted, whereas this behavior is absent in intrinsic AMs.

		\textit{\color{blue}Spin-orbital texture.--}
		To understand the underlying difference among ${\cal O}_{1}$, ${\cal O}_{4}$, and the other phases, we introduce the spin-orbital weight/texture. In Sec.~III of SM~\cite{sm2024}, we derive the effective mean-field ${\bm k}\cdot{\bm p}$ Hamiltonian around the $\Gamma$ (or $M$) point for Eq.~\eqref{eq-mf-ham0}. By projecting out the sublattice degrees of freedom, we obtain 
		\begin{align} \label{eq-kp-gamma}
			\begin{split}
				{\cal H}_{\Gamma} ({\bm k}) = &\; E_0({\bm k}) - [4t_3 k_x k_y \tau_x + 4t_5(k_x^2-k_y^2) \tau_z ] s_0  \\
				&+ [\tilde{\Delta}_1({\bm k}) \tau_0 + \tilde{\Delta}_2({\bm k})\tau_x + \Delta_3\tau_x  + \Delta_4 \tau_z ] s_z,
			\end{split}
		\end{align}
		where $E_0({\bm k})=m_0(k_x^2+k_y^2)$ with $m_0 = 2(t_1 + t_2) - t_0$, $J=(t_1-t_2)/(2t_0)\propto \delta t_{12}$, $\tilde{\Delta}_1({\bm k})=J\Delta_1(k_x^2-k_y^2)$, and $\tilde{\Delta}_2({\bm k})=J\Delta_2(k_x^2-k_y^2)$. This shows the differences between the JT-induced and intrinsic AMs. While the term $\tilde{\Delta}_{1/2}$ are induced by the JT mode, $\Delta_{3/4}$ represent intrinsic AMs. The spin splitting is elucidated from commutation relations, e.g., $[\epsilon_3({\bf k}) \tau_0\sigma_zs_0, \Delta_1\tau_0 \sigma_z s_z ]=0$. This implies that changing the sign of $\delta t_{12}$ is analogous to flipping the sign of $\Delta_1$, which accounts for the observed sign reversal in $\sigma_{xx}^z$. Moreover, the spin-split Fermi surfaces for each AM phase are illustrated in Figs.~\ref{fig2}(a-d), respectively. Specifically, Fig.~\ref{fig2}(b) illustrates the $g$-wave character, while Figs.~\ref{fig2}(a) and \ref{fig2}(c-d) show the $d$-wave AMs, in agreement with our symmetry analysis.

		Remarkably, all AM phases except for ${\cal O}_{1}$ display orbital dependencies, showcasing nontrivial spin-orbital textures. To illustrate this, we first define the spin-resolved orbital weight on each Fermi surface:
		\begin{align}
			p_s^{n}(\theta_k) &= \langle E_{n,s}(k_{f,s}^{{(n)}},\theta_k) | \tau_z | E_{n,s}(k_{f,s}^{{(n)}},\theta_k) \rangle,    
		\end{align}
		where $|E_{n,s}(k,\theta_k)\rangle$ represents the eigenstate of ${\cal H}_\Gamma(k,\theta_k)$ in polar coordinates, $n=\pm$ is the band index without spin-splittings ($\Delta_i=0$), $s=\{\uparrow,\downarrow\}$ denotes spin, and $k_{f,s}^{{(n)}}$ is the associated Fermi momentum. We present the main results in Figs.~\ref{fig2}(e-f) and provide the analytical expressions in Sec.~IV of SM~\cite{sm2024}. Then, we introduce 
		\begin{align}
			\Delta p_s^{n} (\theta_k) = p^{n}_\uparrow(\theta_k) - p^{n}_\downarrow(\theta_k),
		\end{align}
		to capture the spin-orbital texture of the $n$-th band. We show $\Delta p_s^{+} (\theta_k)$ in Fig.~\ref{fig2}(e) and $\Delta p_s^{-} (\theta_k)$ in Fig.~\ref{fig2}(f). Unlike the other phases, $\Delta p_s^\pm$ only vanishes in the ${\cal O}_{1}$ phase~\cite{note-1}. Thus, $\Delta p_s \neq 0$ signifies spin-orbital textures and characterizes
		spin-orbital AM, which features a spin-orbital-locked magnetic ordering. Moreover, $\Delta p_s$ varies significantly among different phases due to their distinct symmetry characteristics, i.e., $\Delta p_s^n(\theta_k) = \Delta p_s^n(\frac{\pi}{2}-\theta_k)=-\Delta p_s^n(\frac{\pi}{2}+\theta_k)$ for ${\cal O}_{2}$, $\Delta p_s^n(\theta_k) = - \Delta p_s^n(\frac{\pi}{2}-\theta_k)$ for ${\cal O}_{3}$, and $\Delta p_s^n(\theta_k) = \Delta p_s^n(\frac{\pi}{2}-\theta_k)$ for ${\cal O}_{4}$.

		Furthermore, the spin-resolved and angle-dependent intensity is defined as 
		\begin{align}
			{\cal I}_s (\theta_k ) = \int_0^\infty k dk \sum_n 
			\delta(E_f-E_{n,s}(k,\theta_k)) p_s^{n}(\theta_k),
		\end{align}
		where $\delta(x)$ is the delta function. The total spin-resolved intensity is then obtained as ${\cal N}_s = \int_0^{2\pi}  {\cal I}_s(\theta_k) d\theta_k$. It measures the orbital polarization ${\cal N}_s = N_{s,d_{xz}} - N_{s,d_{yz}}$, where $N_{s,d_{xz}}$ ($N_{s,d_{yz}}$) denotes spin-resolved density of states in the $d_{xz}$ ($d_{yz}$) orbital. In Figs.~\ref{fig2}(g) and \ref{fig2}(h), we plot $\Delta {\cal N}_s = | {\cal N}_s / (N_{s,d_{xz}} + N_{s,d_{yz}}) |$ as a function of $\Delta_{i}$ and $\delta t_{12}$, respectively. Interestingly, $\Delta {\cal N}_s$ is nonzero for both ${\cal O}_{1}$ and ${\cal O}_{4}$, whereas it vanishes for others. This distinction arises from the sign-changing behavior of $p_s^n$ for the ${\cal O}_{\text{AFM},2/3}$ phases. The dependence on $\delta t_{12}$ can be also used to distinguish ${\cal O}_{1}$ (JT-induced extrinsic AM) from ${\cal O}_{4}$ (intrinsic AM), particularly for $T_N<T_S$.

		\textit{\color{blue}Effect of JT-induced crystal field.--}
		The ${\cal O}_1$ phase lacks a spin-orbital texture. However, this changes when considering a $Q_2$ JT mode. As derived in Sec.~V of SM~\cite{sm2024}, this JT mode introduces both $\delta t_{12}$ and a staggered crystal field, represented as $\delta_{\text{JT}}\tau_z\sigma_z$. When $\delta_{\text{JT}}>J\Delta_1$, a second-order perturbation yields an additional term $\delta_{\text{JT}}\Delta_1/(4t_0)\tau_zs_z$ in Eq.~\eqref{eq-kp-gamma}. This term effectively resembles that of the ${\cal O}_4$ phase, as both phases share the same symmetry classification when $\delta t_{12} \neq 0$. Therefore, the interplay between $\delta_{\text{JT}}$ and ${\cal O}_1$ significantly intensify $\Delta p_s^n(\theta_k)$, which is proportional to $\delta_{\text{JT}}$. We note that $\delta_{\text{JT}}$ can reach substantial magnitudes (e.g., $0.5$ eV as reported in Ref.~\cite{minarro2024spin}), rendering the ${\cal O}_1$-induced $\Delta p_s^n(\theta_k)$ comparable in magnitude to that from the ${\cal O}_2$ phase. We also explore this mechanism for spin-orbital textures in candidate materials (see Sec.~V of SM~\cite{sm2024}). Therefore, we conclude that spin-orbital texture is a universal feature of both intrinsic and extrinsic AMs.

		\begin{figure}[t]
			\centering
			\includegraphics[width=0.95\linewidth]{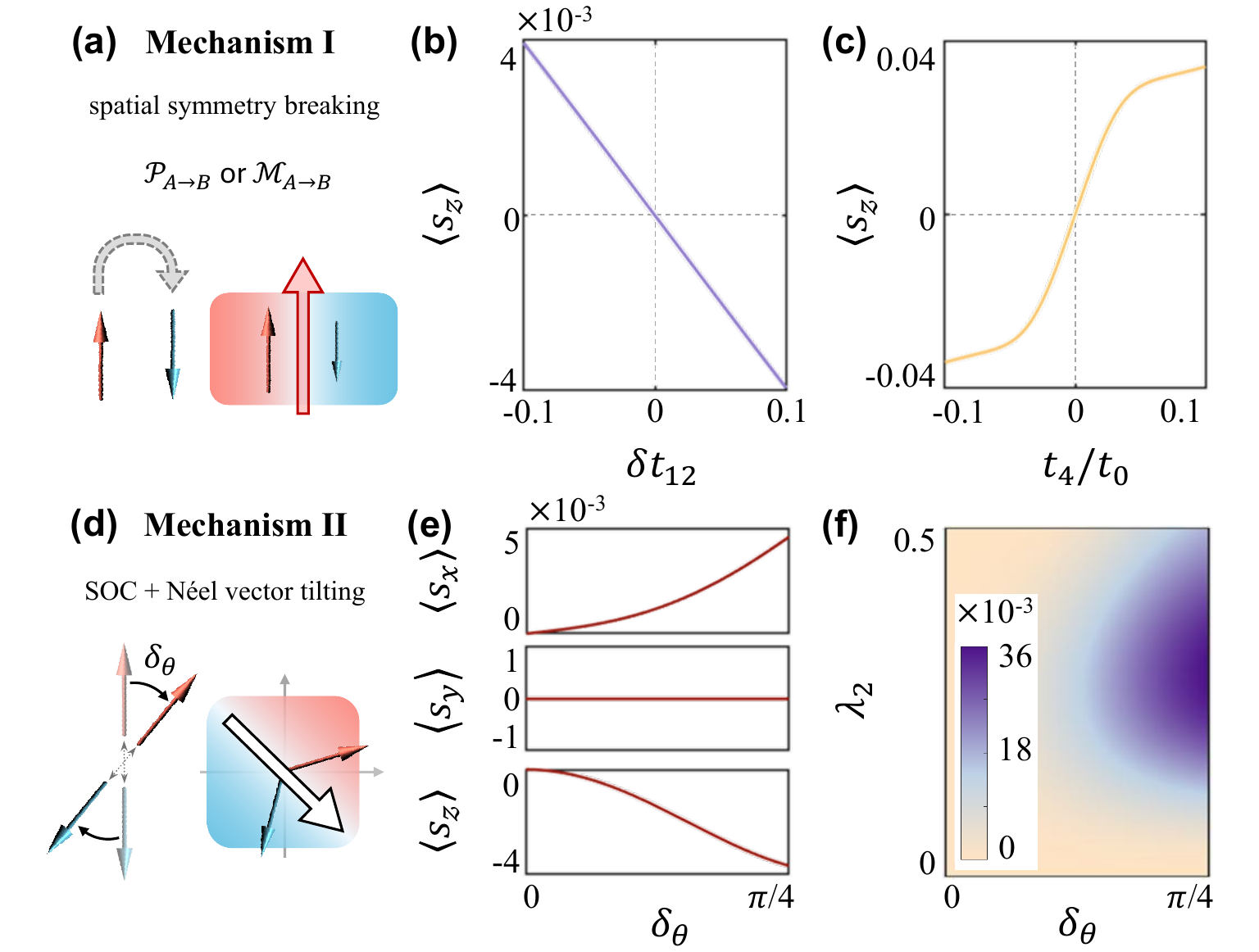}
			\caption{(a) Illustration of weak magnetization (ferrimagnetism) resulting from the breaking of crystalline symmetries in an AM. 
				The calculated $\langle s_z\rangle$ are presented in (b) for ${\cal O}_{5}$ and in (c) for ${\cal O}_{2}$, respectively. 
				(d) Depiction of non-collinear magnetic order generated by SOC, effectively tilting the N\'eel vector.
				Calculations of $\langle \vec{s}\rangle$ for ${\cal O}_{1}$ are shown in (e) and (f). Parameters are the same as those used in Fig.~\ref{fig1}(d).}
			\label{fig3}
		\end{figure}

		\textit{\color{blue}Weak magnetization.--}
		The spin-orbital texture for the ${\cal O}_{5}$ phase are provided in Sec.~VII of SM~\cite{sm2024}. Below, we discuss two general mechanisms for generating magnetization in AMs. The first mechanism arises from the breaking of crystalline symmetries such as ${\cal P}_{A\leftrightarrow B}$ and ${\cal M}_{A\leftrightarrow B}$. Lifting symmetry constraints result in the effective local moments at the two sublattices becoming non-equivalent, leading to ferrimagnetism [Fig.~\ref{fig3}(a)]. We calculate the magnetization $\langle s_z \rangle$ for ${\cal O}_{5}$ as a function of $\delta t_{12}$ [Fig.~\ref{fig3}(b)]~\footnote{Note that in the absence of SOC, the N\'eel vector can point along any arbitrary direction.}. 
		Additionally, an example of ${\cal M}_{A\leftrightarrow B}$-breaking terms, such as $- 2 t_4 (\cos{k_x} - \cos{k_y)} \tau_y\sigma_y s_0$, can induce magnetization in the ${\cal O}_{2}$ phase [Fig.~\ref{fig3}(c)]. In both cases, the system retains $[{\cal C}_2||{\cal C}_{4z}(X)]$, restricting magnetization to align along the N\'eel-vector direction.

		The second mechanism, applicable to both extrinsic and intrinsic AMs, involves SOC. It gives rise to a non-collinear magnetic order consequently generates Berry curvature. In practice, we tilt the N\'eel vector from the $z$ axis to an arbitrary direction (e.g.,~the $x$ axis), such that the four-fold rotation couples spatial rotation with internal spin rotation. Thus, the $[{\cal C}_2||{\cal C}_{4z}]$ symmetry is broken, as illustrated in Fig.~\ref{fig3}(d). The effective polarization angles of the local moments at the two sublattices become misaligned, giving rise to magnetization that plays a role akin to the Dzyaloshinskii–Moriya interaction~\cite{dzyaloshinsky1958thermodynamic,moriya1960anisotropic}, as numerically confirmed in Fig.~\ref{fig3}(e). For the calculations, ${\cal H}_{\text{soc}} = \lambda_1 \tau_y \sigma_0 s_z + \lambda_2 \tau_0 \sigma_z (\sin 2k_x s_y - \sin 2k_y s_x)$ is used, where $\lambda_1$ is the atomic SOC and $\lambda_2$ is the ${\cal P}_{A\leftrightarrow B}$-preserving Rashba SOC. We take the ${\cal O}_{1}$ phase for example and calculate the magnetization as a function of $\lambda_2$ and tilting angle $\delta_\theta$ in Fig.~\ref{fig3}(f). The magnetization intensifies with increasing $\lambda_2$ and $\delta_\theta$. This further generates a hysteresis loop in the anomalous Hall effect~\cite{gonzalez2023prl}, with calculations provided in Sec.~VI of SM~\cite{sm2024}.

		\textit{\color{blue}Possible realization.--}
		Finally, we explore a potential realization of spin-orbital AMs using a mean-field approach and then calculate the staggered susceptibility, $\chi_i(T) = -\tfrac{k_BT}{2} \sum_{i\omega_n,{\bm k}} \text{Tr}[G_0(i\omega_n,{\bm k}) {\cal O}_{i}  G_0(i\omega_n,{\bm k}) {\cal O}_{i}]$, where $T$, $\omega_n$, and $G_0$ are the temperature, Matsubara frequency, and non-interacting Green's function, respectively (see Sec.~VIII of SM~\cite{sm2024}). The dominant contribution to $\chi_i$ arises from inter-band parts, and it converges to a finite value as $T$ approaches zero~\cite{moriya2012spin}. Using the same parameters as in Fig.~\ref{fig1}(d), we plot the maximum value of $\chi_i(T)$ as a function of chemical potential $\mu$ near the half-filling, and find that the JT-induced AMs (${\cal O}_{1/2}$) are predominant [Fig.~\ref{fig4}(a)]. Upon setting $t_5=0$, one can analytically show $\chi_1 = \chi_2$ for the model in Eq.~\eqref{eq-mf-ham0} (see Sec.~VIII of SM~\cite{sm2024}). However, our numerical results show that $\chi_2$ becomes dominant when $t_5\neq0$ [Fig.~\ref{fig4}(b)]. This can be understood in the context of half-filling. For ${\cal O}_1$, the $t_5$ term is inactive in mediating electron hopping within the same sublattice. Whereas, ${\cal O}_2$ permits such hopping, thereby lowering the system's energy.

		We confirm that the momentum-resolved bare susceptibilities $\xi_i(\bm{q})$ exhibit a pronounced peak at the $\Gamma$ point in Sec.~IX of SM~\cite{sm2024}. Therefore, we can calculate the N\'eel temperature by solving the linearized gap equation, $\chi_i(T_N)=1/U_{i}$, where $U_{i}$ represents the effective interaction in ${\cal O}_i$. The absence of divergence in $\chi_i$ suggests a finite critical interaction to stabilize AMs, in agreement with self-consistent calculations. Based on the two-orbital $U$-$U'$-$J_H$ Hubbard model, the effective interactions are $U_{1}=U+J_H$ for ${\cal O}_{1}$ and $U_2=U-J_H$ for ${\cal O}_{2}$. Thus, $U_1$ is always larger than $U_2$ as Hund's rule requires. This indicates that ${\cal O}_2$ is likely the leading phase at smaller $J_H$ but larger $t_5$, giving rise to the spin-orbital AM, as illustrated in the $U$-$J_H$ phase diagram [Fig.~\ref{fig4}(c)]. This is further supported by calculating random-phase-approximation renormalized susceptibilities~\cite{lu2025}. Furthermore, the interaction-driven phase transition from a planar $g$-wave to $d$-wave AM differs from previous reports~\cite{Fernandestoplogical,leeb2024prl,atasi2024prb}. Our theoretical framework for spin-orbital altermagnetism can be generalized to investigate phase transitions between distinct magnetic orders.

		\begin{figure}[t]
			\centering
			\includegraphics[width=0.95\linewidth]{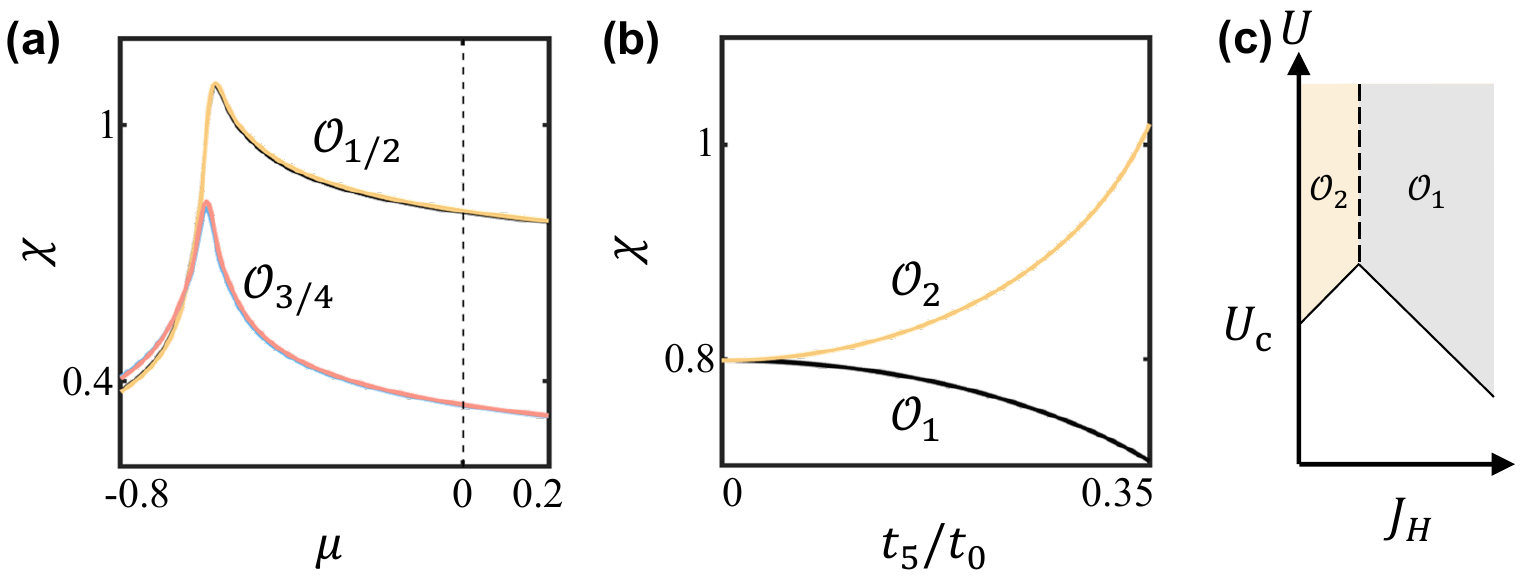}
			\caption{(a) Maximal value of $\chi_i(T)$ as a function of the chemical potential $\mu$ for $\mathcal{O}_{1/2/3/4}$, with the dashed line marking the half-filling. 
				(b) $\chi_1$ and $\chi_2$ (the two leading phases) as a function of $t_5$.
				(c) Schematic $U$-$J_H$ phase diagram for finite $t_5$. Parameters are the same as those used in Fig.~\ref{fig1}(d).}
			\label{fig4}
		\end{figure}

		\textit{\color{blue}Conclusion.--}
		In summary, we show the crucial role of AM phases in achieving non-relativistic spin-orbital texture by considering the intertwined orderings between spin and orbital degrees of freedom. We demonstrate that both extrinsic (e.g.,~JT-induced) and intrinsic AMs can host nontrivial spin-orbital textures with distinct characteristics. These AM phases can be detected via spin conductivity and spin-resolved orbital polarization for a two-orbital interacting model. We also discuss potential realizations for spin-orbital AMs based on the Ginzburg-Landau theory. Additionally, weak magnetization induced by symmetry breaking, with or without SOC, can occur in the AMs. Our results suggest that other structural phase transitions beyond JT distortions can also lead to the crossover from a N\'eel antiferromagnet to an AM. Thus, our work can guide future experiments on searching materials and observing those intriguing states, particularly in square-planner systems with $d^1$ or $d^2$ high-spin and $3d$ low-spin states, as well as in octahedral systems with $d^1$ or $d^2$ high-spin and $d^3$ or $d^4$ low-spin states. The spin–momentum–orbital locking in spin-orbital altermagnets may enable efficient electrical magnetization control by spin–orbit torques.

		\textit{\color{blue}Acknowledgments.--}
		We thank K.-J.~Yang, H.-M.~Yi, M.~Zeng, and C.-X.~Liu for helpful discussions. We thank C.~Lu for discussions and collaborations on closely related projects.
		L.H.H. is supported by National Key R\&D Program of China (Grant No. 2022YFA1402200), the National Natural Science Foundation of China (Grants No. 12034017).
		Z.M.W. and D.H.X. were supported in part by the NSFC (Grant Nos.~12074108, 12474151 and 12347101), the Natural Science Foundation of Chongqing (Grant No.~CSTB2022NSCQ-MSX0568).
		S.B.Z. acknowledges the support of the start-up fund at HFNL, the Innovation Program for Quantum Science and Technology (Grant No.~2021ZD0302800), and Anhui Initiative in Quantum Information Technologies (Grant No.~AHY170000). 
		Z.M.W. and L.H.H. is supported by the start-up of Zhejiang University and the Fundamental Research Funds for the Central Universities (Grant No. 226-2024-00068).

		\textit{Note added.--} After the submission of our manuscript, we became aware of of several recent preprints addressing related phenomena, including spin-orbital locking effects in extrinsic altermagnets (${\cal O}_1$)~\cite{vila2024orbital}, and candidate materials for intrinsic altermagnetism (${\cal O}_4$)~\cite{jaeschke2025atomic,d2025altermagnetism}.
		
		\vspace{2pt}
		\noindent 
		\textit{Author contributions:}
		L.H.H. initiated the idea and led the project. Y.Z., S.B.Z., and D.H.X. contributed to the development of the idea. Z.M.W., Y.Z., and L.H.H. performed the calculations with assistance from S.B.Z. and D.H.X. L.H.H., Y.Z., and S.B.Z. wrote the manuscript with input from all authors.
		Z.M.W. and Y.Z. contributed equally to this work.
		
		\bibliography{ref}
	\end{document}